\documentstyle[12pt,a4,epsfig]{article}
\newcommand{\be}{\begin{equation}}
\newcommand{\ee}{\end{equation}}
\newcommand{\ba}{\begin{eqnarray}}
\newcommand{\ea}{\end{eqnarray}}

\newcommand{\halfs}[1]{\raisebox{-3mm}[0pt][0pt]{#1}}
\newcommand{\e}{\mbox{e}}

\newcommand{\figcaption}{\refstepcounter{figure}\par\noindent
{Figure \arabic{figure}~ }}
\newcommand{\pp}{{\it pp} }
\begin{document}

\hfill SPhT-98/145

\hfill FIAN/TD-28/98

24/Jan/2000

\bigskip

\bigskip

\begin{center}
{\bf ANGULAR PATTERN OF MINIJET TRANSVERSE ENERGY FLOW IN HADRON AND NUCLEAR
COLLISIONS}
\end{center}
\medskip
\begin{center}
{\bf Andrei Leonidov$^{(a,b)}$ and Dmitry Ostrovsky$^{(a)}$}
\end{center}

\medskip
\begin{center}
{\it (a) P.N.~Lebedev Physics Institute, 117924 Leninsky pr. 53,\\
 Moscow, Russia}
\end{center}
\begin{center}
{\it (b) Service de Physique Theorique, C.E. Saclay, F-91191\\
Gif-sur-Yvette, France}
\end{center}
\bigskip
\begin{center}
\bf Abstract
\end{center}

 The azimuthal asymmetry of minijet system produced at the early stage of
nucleon-nucleon and nuclear collisions in a central rapidity window
is studied. We show that in \pp collisions the minijet transverse energy
production in a central rapidity window is essentially unbalanced in azimuth
due to asymmetric contributions in which only one minijet hits the
acceptance window. We further study the angular pattern of transverse energy
flow generated by semihard degrees of freedom at the early stage of high energy
nuclear collisions and its dependence on the number of semihard collisions in
the models both including and neglecting soft contributions to the inelastic
cross section at RHIC and LHC energies as well as on the choice of the infrared
cutoff.

\newpage

\section{Introduction}

Minijet physics is one of the most promising applications of perturbative QCD
to the analysis of processes with multiparticle production. It addresses a
crucial question of how many (semi)hard degrees of freedom can be available in
a given event. The approach is based on the idea that some portion of
transverse energy is produced in the semihard form, i.e. is perturbatively
calculable because of the relatively large transverse momenta involved in the
scattering, but, due to parametrically strong hadronization effects,
 can not observed in the form of customary well collimated hard jets distinctly separable from the soft background  This mechanism operates at
the early stage of the collision and, when relevant, determines the
characteristics of the primordial transverse energy flow.

Creation of many (semi)hard degrees of freedom corresponds to a new physical
situation characterized by nontrivial, possibly kinetic or even hydrodynamic,
phenomena occurring at the parton level at the early stages of high energy
collision. Of special relevance are here the ultrarelativistic heavy ion
collisions, where one would expect that creating a dense system
of (semi)hard degrees of freedom in the volume much larger than, e.g., the
proton one, possible, thus making the application of concepts borrowed from
macroscopic physics natural. Recent critical discussion of this field can be
found in \cite{GG}.

Minijet physics is an actively developing field.  Reviews on the subject
containing a large number of references are,  e.g., \cite{LR}, \cite{XNW} and
\cite{KE1}. Several approaches have been considered with an aim of providing a
quantitative description of primordial parton system produced at the earliest
stage of, e.g., high energy heavy ion collisions. The conceptually simplest one
is based on the standard  formalism of collinearly factorized QCD at small
parton densities, see \cite{KLL}, \cite{EKL}, \cite{EKR1}, \cite{EK},
\cite{EKR2}. Here one operates with a single hard parton-parton scattering in a
given hadron-hadron collision, so that standard QCD structure
functions can be used in  computing the probability of generating a pair of partons with certain kinematical characteristics.

This approach has a natural generalization, in which multiple
binary parton-parton collisions in the given hadron-hadron one are considered,
provided some ad hoc distribution in the number of these collisions is chosen,
see e.g. \cite{SZ}. This also allows to construct geometrically  motivated
scheme of unitarizating the semihard contribution to the inelastic cross
section of hadron scattering as described, e.g., in \cite{XNW}.

Starting from \cite{BM}, nonlinear QCD effects in relation with describing the
early stages of heavy ion collisions drew progressively more and more
attention. New  results were obtained within the  approach to minijet
production based on the quasiclassical treatment of nuclear gluon distributions
within a framework of McLerran-Venugopalan model \cite{MV}, see \cite{KMW},
\cite{KR}, \cite{MMR}, \cite{GM} and \cite {KM}. First nonperturbative results
on gluon production are now available \cite{KV}, see also \cite{BMP} and
\cite{MP}. Recently a nonperturbative model for gluon production in heavy ion
collisions based on the physical concepts of McLerran-Venugopalan model and a
corresponding kinetic equation describing the evolution of primordial gluon
system were discussed in \cite{AM}.  For  pedagogical introduction to this
rapidly developing field see the lectures \cite{LM} and \cite{AM1}.

A notable feature of the physical phenomena related to the collective behavior
of multiparton systems is their  genuine event-by-event nature, so that many
usual tools used in high energy physics, such as inclusive distributions, are
becoming less helpful.  Thus the analysis of event-by-event variations of the
quantities sensitive to the collective dynamics is very important, see. e.g.
\cite{St}, \cite{SH}, \cite{GRL}, \cite{SRS} and references therein.

The description of the primordial parton configuration should provide
information on the event-by-event pattern of the parton system, in particular
on the number of perturbative parton-producing interactions, which, to a large extent, determines the initial parton density and other kinematical characteristics.
In particular, the discrete nature of parton production in phase space, as
described by finite order QCD calculations, can give rise to primordial
event-by-event angular asymmetries of the parton flow. The fate of the
primordial angular asymmetries depends on the relevant dynamics (is the
evolution of the produced system long enough to wash them out, can they be
frozen and directly relevant to the observed hadronic flow, etc). In any case
the first problem to look at is to study the primordial parton system before
the reinteraction of partons sets in.

   The aim of this paper is to study the characteristics of the initial
minijet-induced transverse energy flow in nucleon-nucleon and nucleus-nucleus
collisions within the framework of minijet production scenario based on
collinearly factorized QCD, \cite{KLL} and \cite{EKL}.  In particular, we shall
 analyze the fluctuational azimuthal imbalance in the minijet transverse energy
flow due to a discrete nature of transverse energy production through basic QCD
hard scattering. We can expect that the effect will be essentially sensitive to
the number of semihard scatterings. In what follows we shall see that this is
indeed the case.

  Below we study the event-by-event inhomogenities in the azimuthal
distribution of minijets following from the basic asymmetry of minijet
transverse energy production into a finite rapidity window in $pp$ collisions.
The nuclear collisions are described by a geometric model \cite{EKL} in which
they are considered  as a superposition of basic nucleon-nucleon ones.  The
azimuthal asymmetry of the minijet system will be specified in terms of
(transverse) momenta only, as calculated in the conventional S-matrix field
theory formalism without referring to the coordinates of partons and making no
assumptions on the structure of the contributions of higher order in QCD
coupling constant in describing the transverse energy production in the
elementary nucleon-nucleon collision. This analysis can be extended to the
next-to-leading order (e.g.  along the lines of \cite{LO}) due to the infrared
stability of the considered distributions which are of energy-energy correlation type.

 The analysis of the event-by-event pattern of the initial minijet generated
transverse energy flow was first presented in \cite{GRZ}, where a HIJING
\cite{HIJING} generated list of partons with specified coordinates and momenta
was used to compute a coarse - grained energy density and velocity field at
RHIC energy $\sqrt{s}=200$ GeV. The resulting distributions turned out to be
highly irregular and similar to the ones occurring in turbulent flows.  Note
that  besides the parton-parton scattering described by collinearly factorized
QCD considered in the present paper, HIJING makes explicit assumptions on the
structure of higher order contributions (unitarization), contribution from
initial and final state radiation and structure of parton system in coordinate
space (thus going beyond standard S-matrix formalism). The existence of
asymmetry due to imbalanced particle production from minijets into a finite
acceptance was mentioned in \cite{XNW}.

The outline of the paper is as follows.

 In the second section we analyse the basic mechanism for producing azimuthally
symmetric and asymmetric configurations in the restricted phase space domain,
which in the considered case will be a unit central rapidity window, in $pp$
collisions.  We calculate in the leading twist
(lowest order in parton density) and
leading order collinear factorization scheme the relative weights for symmetric
(two-jet) and asymmetric (one-jet) contributions to the transverse energy
production cross section for RHIC ($\sqrt{s} = 200$ GeV) and LHC ($\sqrt{s} =
5500$ GeV) for underlying nucleon-nucleon collisions.

 In the third section the computed  contributions to azimuthally symmetric and
asymmetric components of the $pp$ minijet transverse energy production into a
unit central rapidity window are used in calculating  the asymmetry of
transverse energy production in heavy ion collisions, where the nuclear
collision is described as a superposition of the nucleon-nucleon ones in a
geometrical approach of \cite{EKL}. We study the azimuthal asymmetry for RHIC
and LHC energies for central collisions for two dynamical
scenarios. In the first scenario the transverse energy production is assumed to
occur through two physically different mechanisms, the soft one and the
(semi)hard one. As our aim is to study the transverse energy flow at early
collisions stage related to the semihard degrees of freedom, the contribution
of soft interactions will be  accounted for only in determining the relative
yield of semihard contribution.  In the second scenario, which can become
 realistic at LHC energies, all primordial transverse energy production is
assumed to occur through the semihard mechanism.

In the last section  we discuss the  results and formulate the conclusions.

\section{Azimuthal pattern of minijet production in \pp collisions}

  The mechanism responsible for transverse energy production in the leading
order in QCD perturbation theory is an elastic two-to-two parton-parton
scattering. Its cross section is given by the standard collinearly factorized
expression
\be
\frac{d\sigma}{dp_{\perp}^2dy_1dy_2}
\, = \, x_1\, f(x_1,p_{\perp}^2)\, \frac{d\hat{\sigma}}{dp_{\perp}^2}\,x_2\,
f(x_2,p_{\perp}^2),
\label{cross}
\ee
where $xf(x,p_{\perp}^2)$ is a parton structure function,
$x_{1,2}=p_{\perp} (e^{\pm y_1}+e^{\pm y_2})/\sqrt{S}$ are the
fractional longitudinal momenta of the produced partons and
$d\hat{\sigma}/dp_{\perp}^2$ is a differential cross section of elastic
parton-parton scattering. In the following we will be specifically interested
in the transverse energy production into some given (central) rapidity interval
$y_{min}<y_1,y_2<y_{max}$.  Operationally the transverse energy $E_{\perp}$
deposited in this window by the two scattered partons is defined as \footnote{In
Eq.~(\ref{tren}) and below $p_i=|{\bf{p}}_{\perp i}|$}
\be
E_\perp \, =  \, p_1\, \theta(y_{min} \le y_1 \le y_{max})+
p_2\, \theta(y_{min} \le y_2 \le y_{max})
\label{tren}
\ee
In the following we shall confine ourselves to considering the central rapidity
interval $y_{min}=-0.5 < y < y_{max}=0.5$ and stay at the LO (Born elastic
scattering) level, so that in each collision the transverse momenta of the two
produced partons are equal,\ $p_{\perp 1}=p_{\perp 2}=p$. This {\it does not}
mean that these transverse momenta will be balanced in the rapidity window
under consideration, so the event space for transverse energy deposition can be
summarized by
\be\label{ev_space}
E_\perp=\left\{
\begin{tabular}{rl}
0 & if no particle gets into the gap\\
p & if one particle gets into the gap\\
2p & if two particles get into the gap\\
\end{tabular}
\right.
\label{evsp}
\ee
When considering the transverse energy production into a given rapidity window
in \pp collisions only the second and third possibilities are relevant.
To  quantify the computation of the contribution corresponding to cases 2 and
3 in (\ref{evsp}) it is convenient to introduce the integral operators
\ba
S_1&=&\int dy_1 dy_2\,
(\theta(y_1)+\theta(y_2)-2\theta(y_1)\theta(y_2))*(\dots)\\
S_2&=&\int dy_1 dy_2\, (\theta(y_1)\theta(y_2))*(\dots)\ ,
\ea
where $\theta (y_{1,2}) =\theta(y_{min}<y_{1,2}<y_{max})$
Applying these operators to the differential cross section Eq.~(\ref{cross}) we
get a decomposition  of the transverse energy production cross section in a
given rapidity window into the separate one-jet and two-jet contributions
(second and third entries in the event list Eq.~(\ref{evsp}))
\be
\frac{d\sigma}{dE_\perp} \, = \, \frac{d\sigma_1}{dE_\perp}+
\frac{d\sigma_2}{dE_\perp},
\label{dec}
\ee
where
\be
\frac{d\sigma_1}{dE_\perp} \,= \,
 S_1*\left( \frac{d\sigma}{dp} \right) \bigg|_{p=E_\perp},
\label{one}
\ee
and
\be
\frac{d\sigma_2}{dE_\perp} \, = \,
 2\,  S_2*\left( \frac{d\sigma}{dp} \right) \bigg|_{p=2E_\perp}
\label{two}
\ee
On the event-by-event basis these contributions correspond to completely
distinct possibilities of having the azimuthally balanced symmetric or
unbalanced asymmetric transverse energy flow in the rapidity window under
consideration.

In Fig.~(1) and Fig.~(2)
\begin{figure}[p]
 \begin{center}
 \epsfig{file=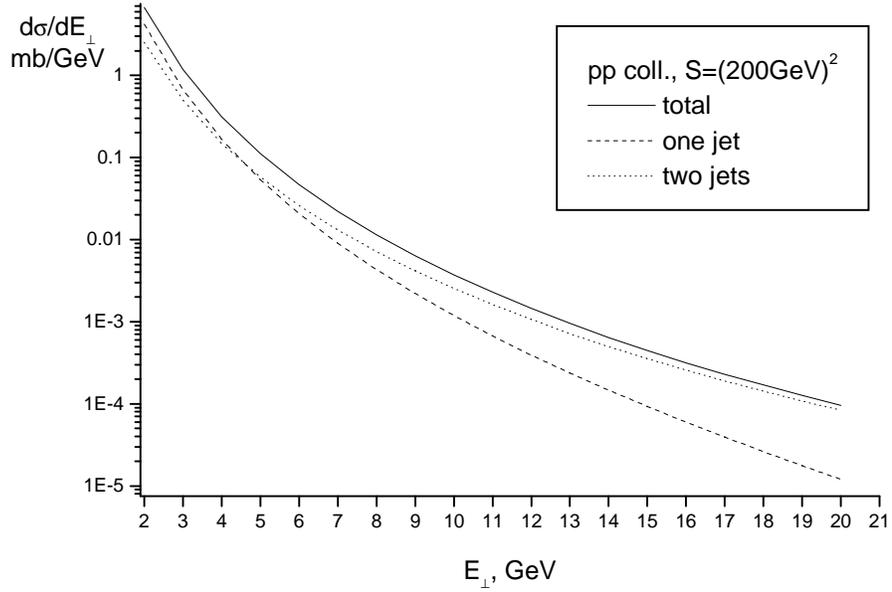,height=8cm}
 \end{center}
 \figcaption{One and two jets contributions to transverse energy production
 in $pp$ collisions in unit central rapidity window at RHIC energy
 $\sqrt{s} = 200$ GeV.}
 \label{200onetwo}
\end{figure}
\begin{figure}[p]
 \begin{center}
 \epsfig{file=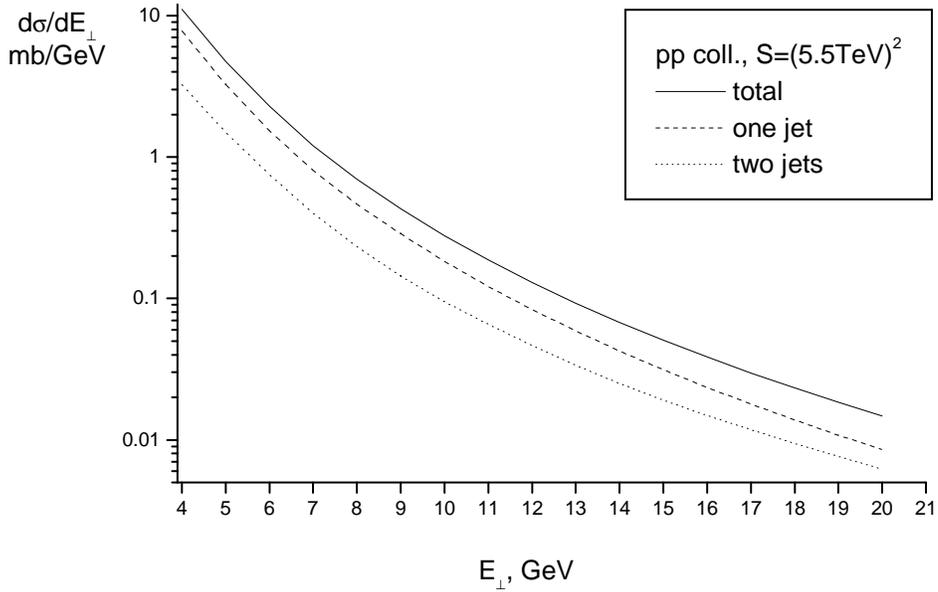,height=8cm}
 \end{center}
\figcaption{One and two jets  contributions to transverse energy production  in
$pp$ collisions in unit central rapidity window at LHC energy $\sqrt{s} = 5500$
GeV.}
 \label{5500onetwo}
\end{figure}
we plot the transverse energy production
cross sections
Eq.~(\ref{one}) and Eq.~(\ref{two}) for RHIC and LHC energies
$\sqrt{s}=200$ GeV and $\sqrt{s} = 5500$ GeV, where for LHC we have chosen
the energy to be available for protons in lead beams and the MRSG structure
functions \cite{MRSG} were used.

\begin{figure}[h]
\begin{center}
\epsfig{file=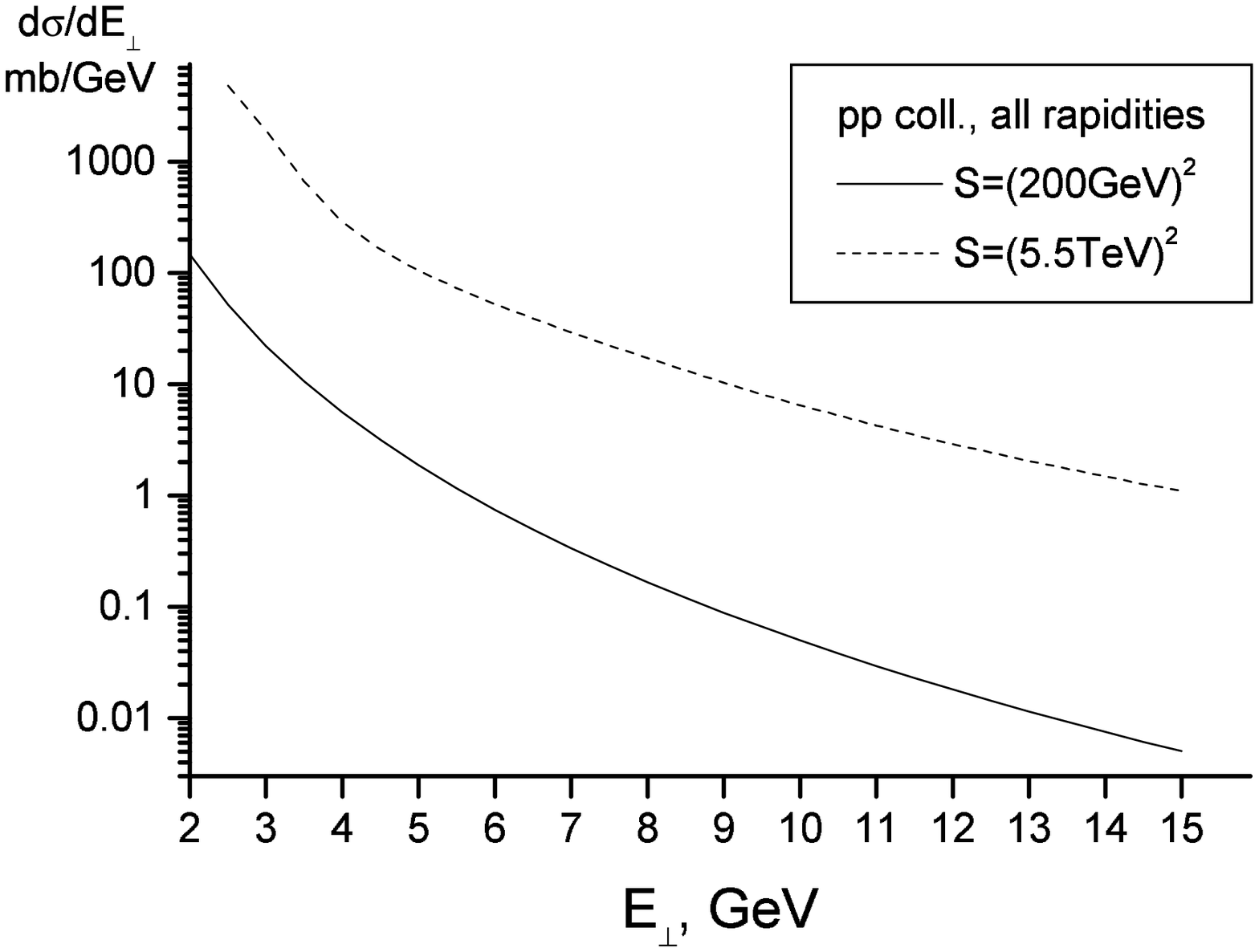,height=8cm}
 \end{center}
\figcaption{Transverse energy production in $pp$ collisions in the whole rapidity
interval at RHIC ($\sqrt{s} = 200$ GeV) and LHC ($\sqrt{s} = 5.5$ TeV)
energies.}
 \label{total}
 \end{figure}

In Fig.(\ref{total}) we also present the differential cross sections of
transverse energy production into full rapidity intervals available at RHIC and
LHC energies which will be used in the next section to normalize
the differential cross sections of transverse energy
production in $pp$ collisions.

 The information contained in Fig.~(1) and (2) is summarized in Table~1, where
we show the parameters for the fits for the one-jet and two-jet spectra
Eq.~(\ref{one}) and Eq.~(\ref{two}) having the functional form
$c\,(E_{\perp}/1\, \mbox{GeV})^{- \alpha}$.

\medskip

\medskip

\begin{center}
\begin{tabular}{|l|c|c|r|r|}
\hline
\halfs{$\sqrt{S}$,GeV} &{ $\alpha_1$}&{$\alpha_2$} &%
\multicolumn{2}{|c|}{ c, mb/GeV}\\
\cline{2-5}
&1 jet& 2 jets&1 jet&2 jets\\
\hline
200             & 5.09 & 4.53 &  173 &  77\\
\hline
5500            & 4.25 & 3.93 & 3099 & 819\\
\hline
\end{tabular}
\bigskip

{\bf Table~1}
\end{center}

\medskip

We see that at RHIC energies the one-jet asymmetric contribution dominates at
small transverse energies, and at $E_{\perp} \sim 4.5$ GeV the two-jet
symmetric contribution takes over. At LHC energies the asymmetric contribution
is clearly dominant in the whole minijet transverse energy range.

\section{Azimuthal asymmetry of minijet transverse energy flow in nuclear
collisions}.

In this section we turn to the analysis of the angular asymmetry of minijet
produced transverse energy flow in  nuclear collisions induced by the
fundamental asymmetry in the \pp collisions described in the previous section.
The translation of the features characterizing the particle production in \pp
collisions to those characterizing the nuclear ones is possible, e.g., in a
geometrical model, where the nucleus-nucleus collision is considered as a
superposition of the proton-proton ones, see e.g.  \cite{KLL}, \cite{EKL}.  At
 each impact parameter $b$, where $b$ is a distance in the transverse plane
between the centers of the colliding nuclei, the nucleus-nucleus collision is
described as a Poissonian superposition of the nucleon-nucleon collisions such
that a probability of $n$ \pp collisions is given by
\be
w_n(b) \, = \, \frac{1}{n!}\, \bar{N}^n_{AB}(b)\, \e^{-\bar{N}_{AB}(b)},
\label{Poisson}
\ee
where $\bar{N}_{AB}(b)$ is the average number of \pp collisions in the
nucleus-nucleus one, which is thus described as a specific superposition of
multiple independent pp collisions occurring with the weight given by
Eq.~(\ref{Poisson}). The elementary $pp$ collisions occur between some nucleon
belonging to the nucleus A located at the transverse distance $b_1$ from its
center with the probability given by the probability density $\rho_A(b_1)$ with
the normalization
\be
\int d^2b_1\, \rho_A(b_1) \, = \, 1.
\ee
and the nucleon from nucleus B  located at the transverse distance $b_2$ from
its center with the probability given by the probability density $\rho_B(b_2)$
with the collision probability
 $P \left(\bar{b}-\bar{b}_1+\bar{b}_2 \right)$. Thus the
average number of \pp collisions characterizing the basic Poissonian process
Eq.~(\ref{Poisson}) is given by
\be
\bar{N}_{AB}(b)\, = \, AB\int d^2b_1d^2b_2\,
 P \left( \bar{b}-\bar{b}_1+\bar{b}_2 \right)\,
\rho_A(b_1)\, \rho_B(b_2),
\ee
where $P(b)$ is a probability of an inelastic collision of two nucleons
initially separated by the transverse distance $\bar{b}-\bar{b}_1+\bar{b}_2$.
The physical meaning of the collision probability $P(b)$ depends on the
underlying physical mechanism responsible for inelastic transverse energy
production in the binary nucleon-nucleon collisions. Our discussion is
confined to minijets as providing such a source, so in our case $P(b)$ is a
probability of minijet producing inelastic nucleon-nucleon collision occurring at fixed impact
parameter $b$. Let us stress that the
differential probability of minijet-induced transverse energy  production
depends on the rapidity window under consideration. The usual
assumption about the impact parameter dependence of the probability of
nucleon-nucleon collisions $P(b)$ is that the collisions are local in the impact parameter plane, i.e.
\be
P(b) \, = \, \sigma^{minijet}_{pp}\,(|\Delta y | \le y_0)\,\delta^{(2)}(b).
\label{distr}
\ee
Let us now discuss in some details the normalization of the Poisson process
Eq.~(\ref{Poisson}) provided by the overall minijet production cross section
into the rapidity window $|\Delta y | \le y_0$,
$\sigma^{minijet}_{pp}(|\Delta y| \le y_0)$.
The overall minijet contribution to transverse energy production
cross section is given by the integral over $E_{\perp}$ of the differential
cross section Eq.~(\ref{dec}). Because of the singular behavior of the
perturbative transverse energy production cross section at small $E_{\perp}$
the very definition of the overall contribution of minijet mechanism to
transverse energy production requires introducing a cutoff at small
transverse energies
\be
\sigma^{minijet}_{pp}\, (|\Delta y | \le y_0|E_0) \,= \,\int_{E_0}
d E_{\perp}\, {d \sigma \over dE_{\perp}} (\Delta y)
\label{norm1}
\ee

Let us note that for any rapidity interval
\be
\int_{E_0} d E_{\perp}\, {d \sigma \over dE_{\perp}}\, (\Delta y) \,  \leq
\, \sigma^{inel} (\Delta y),
\label{norm2}
\ee
where $\sigma^{inel} (\Delta y)$ is an (experimental) inelastic
cross section in
a given rapidity window $\Delta y$. This shows that the cutoff $E_0$ is
physically a function of the considered rapidity interval.

 Another important issue related to the choice of this cutoff is the possible
contribution to the overall inelastic cross section of other mechanisms of
transverse energy production, e.g. of the soft particle production due to the
decay of hadronic strings. The restriction Eq.~(\ref{norm2}) clearly refers
only to the part of inelastic cross section corresponding to hard inelasticity,
i.e. transverse energy production through semihard processes. The other part of
the inelastic cross section corresponds to soft mechanisms of transverse energy
production which do not involve large momentum transfers. It is important to
note, that the characteristic time scale of semihard transverse energy
production are smaller than that for the soft nonperturbative mechanism. At
early stages of the collision hard parton skeleton is formed, which is then
dressed by soft particle production due to strings stretching in between the
partons originating from primordial processes characterized by large momentum
transfer. This shows, in particular, that the soft processes do not have,
generally speaking, an independent share of the overall inelasticity, so  the
naive additivity
\be
{d \sigma \over d E_{\perp}} \, = \,
{d \sigma^{minijet} \over d E_{\perp}} + {d \sigma^{soft} \over d E_{\perp}}
\ee
is, in general, not valid. It could happen, e.g., that  with growing
collision energy the yield of events with hard initial inelasticity would
be dominant or even cover the whole event space (here we refer to
non-diffractive contribution). In such extreme scenario the {\it only}
function of soft mechanism is stretching the strings between the {\it hard}
initial partons. Here it is important to recall that the
cross section of transverse energy
production Eq.~(\ref{dec}), as computed in perturbative QCD, is a so-called
infrared safe quantity and is thus entirely determined by its early
quark-gluon stage and does not depend on the late stages of the process,
including string formation between the separating partons.

Let us stress once again, that in the present study we confine our
consideration to analyzing the angular pattern of the primordial transverse
energy flow generated at the early stages of the collisions by the semihard
degrees of freedom (minijets). The analysis of the effects related to the
subsequent redistribution of primordial transverse energy by soft interactions
at larger times will be discussed in the future publications \cite{LO1}.

 The yield of perturbative contribution as a function of CMS energy is
a crucial characteristics of the inelastic cross section. Unfortunately
very little can currently be said about its magnitude, resulting in the
uncertainty in fixing the cutoff for the perturbative contribution.

In view of this we shall fix the cutoff value $E_0$ at given collision energy
as follows. To explore the possible "window of opportunities" for the hard
minijet contribution as determined by the yield of independent soft particle
production we will discuss two model scenarios. Namely, when considering the
inelastic particle production at all rapidities we shall either assume the
constant soft contribution $\sigma^{soft}(pp) = 32 $\ mb, the inelastic cross
section of $pp$ scattering at intermediate energies, universally present at all
CMS energies (mixed scenario) or assume that in all collisions the
transverse energy is produced via the early perturbative minijet stage, i.e.
put $\sigma^{soft}(pp) = 0$ (hard scenario). The cutoff $E_0$ is thus
determined from \footnote{The inelastic cross section is computed using the
parameterization
$\sigma_{inel}(s)=\sigma_0\cdot(s/s_0)^{0.0845}\cdot(0.96-0.03\cdot\log
(s/s_0))$, where $s_0=1$ GeV, $\sigma_0=21.4$ mb, which gives a good
description of the existing experimental data \cite{exp}, see also a
compilation in \cite{XNW}}
\be
\int_{E_0} d E_{\perp} {d \sigma^{minijet}_{pp} \over dE_{\perp}} =
\sigma^{hard}=
\left\{
\begin{tabular}{ll}
$\sigma^{inel}_{exp},$ & no soft contribution\\
$\sigma^{inel}_{exp}$\, -- \, 32\, mb, & soft contribution 32\, mb,\\
\end{tabular}
\right.
\label{norm3}
\ee
where the transverse spectrum in Eq.~(\ref{norm3}) refers to the full
kinematic interval (see Fig.~(\ref{total})). Let us stress that the
differential cross section for transverse energy production that we
use in Eq.~(\ref{norm3}) is the result of the lowest order calculation
from the previous section, and the higher order effects that can
phenomenologically be included within the geometrical unitarization scheme,
see e.g. \cite{XNW}, are not included.
Numerical values of the cutoff found by integrating the spectra shown in
Fig. (\ref{total}) are given in Table 2

\medskip

\begin{center}
\begin{tabular}{|c|c|c|c|c|c|c|}
\hline
$\sqrt{S},GeV$& $\sigma^{soft}$, mb & $\sigma^{hard}, mb$ & %
$E_0$, GeV& $p_1$ & $\sigma_{pp}^{minijet}$, mb&%
$\sigma_{PbPb}^{minijet}$, mb\\
\hline
\halfs{200} & 0&41.8 &2.4&0.54&2.4&5336\\
\cline{2-7}
           & 32 & 9.8 &3.5&0.48&0.54&4102\\
\hline
\halfs{5500}& 0 &66.3 &6.9&0.65&2.8&5443\\
\cline{2-7}
           & 32 &34.3 &8.4&0.64&1.5&4970\\
\hline
\end{tabular}

\medskip
{\bf Table 2}
\end{center}

In the fifth column we show the overall probability
of asymmetric one-jet contribution $p_1(E_0)$ calculated using the
differential transverse energy production spectra Eq.~(\ref{dec}),
(\ref{one}) and (\ref{two}):
\be
p_1(E_0) \, = \,
 \left.\left(
 \int_{E_0}^{\infty} dE_{\perp}\, {d \sigma_1 \over dE_{\perp}}
 \right) \right/
 \left(
 \int_{E_0}^{\infty} dE_{\perp}\, {d \sigma \over dE_{\perp}}
 \right).
\label{p1}
\ee
As already mentioned, although the differential spectra describe the
transverse energy production into some given rapidity window,
in what follows  the value of the cutoff $E_0$ will be determined
from Eq.~(\ref{norm2}) considered for full rapidity window kinematically available for inelastic energy production at given
collision energy. For more accurate determination of the cutoff
$E_0$ one would need  experimental data on inelastic cross sections
in, e.g., central rapidity window.  Different quantities have different
sensitivity with respect to the choice of the cutoff and, in particular,
the dependence of $p_1(E_0)$ in Eq.~(\ref{p1}) on $E_0$ is quite weak.
In the last column in Table~2 we show the cross section of producing
at least one minijet in lead-lead collisions $\sigma^{minijet}_{PbPb}$:
\be
\sigma^{minijet}_{PbPb}=\int d^2b\, (1-w_0(b))\, = \,
 \int d^2b\, (1-\e^{-\bar{N}_{AB}(b)}).
\label{spb}
\ee
where $w_0(b)$ is a probability of having no minijet producing
nucleon -- nucleon collisions, cf. Eq.~(\ref{Poisson}).

The transverse energy production in nucleus-nucleus collisions is then
described by the convolution of the distribution over the number of $pp$
collisions
obtained from Eq.~(\ref{Poisson}) at given impact parameter with
the distributions characterizing the transverse energy production in $pp$
collisions Eq.~(\ref{one}) and Eq.~(\ref{two}).

In practice this convolution was realized by a Monte Carlo procedure, where
\begin{itemize}
\item a large number ($10^7$) nucleus-nucleus collisions were generated
with the number of $pp$ collisions $N$ distributed according to
Eq.~(\ref{Poisson}),
\item the weight of one-jet asymmetric (two-jet symmetric) $pp$
collisions is equal to the probability $p_1$ (respectively $1-p_1$) with
$p_1$ taken from Table~2. More explicitly, this corresponds to a binomial
distribution in the number of asymmetric collisions $N_a$
\be\label{binom}
w(N_a) = C^{N_a}_{N}\,p_1^{N_a}\,(1-p_1)^{N-N_a}
\ee
\item the weight for $E_{\perp}$ itself was in turn determined by
Eq.~(\ref{one}) and Eq.~(\ref{two}) for asymmetric and symmetric contributions
correspondingly,
\item the azimuthal orientation of jet(s) was determined
at random corresponding to a flat distribution in the azimuthal angle.
For two-jet events the jets are going into opposite directions,
so that their azimuth differs by $\pi$.
\end{itemize}

Let us now turn to the quantitative analysis of the  event-by-event asymmetry
of the minijet generated transverse energy flow. Our analysis will be made
using a (normalized) difference between the transverse energy flow into the
oppositely azimuthally oriented sectors with a specified angular opening
$\delta \varphi$ each and rapidity window $|y|<0.5$. Let us note that this
quantity has an important advantage of allowing for the future next-to-leading
order analysis.  For convenience one can think of the directions of these cones
as being  "up" and "down" corresponding to some specific choice of the
orientation of the system of coordinates in the transverse plane. All our
results are, of course, insensitive to the particular choice.  Let us denote the
transverse energy going into the "upper" and "lower" cones in a given event by
$E_\uparrow (\delta \varphi)$ and $E_\downarrow(\delta \varphi)$
correspondingly. The magnitude of the asymmetry in transverse energy production
can then be quantified by introducing a variable
\be\label{deltaE}
\delta E \, = \, E_\uparrow (\delta \varphi) - E_\downarrow (\delta \varphi)
\ee
Using the distribution over the number of asymmetric collisions
Eq.~(\ref{binom}) and taking into account that the event space of asymmetric
collisions is further subdivided into two sets corresponding to nonzero
energy going into the upper and lower cone $E_{\uparrow}$ and $E_{\downarrow}$,
we can calculate the quadratic mean of $\delta E (\delta \phi)$
for the considered azimuthal openings $\delta \phi =
\pi/2^n$\,\,$(n=0,1,2)$
\footnote{It is easy to see that the distribution of $\delta E$ is a so-called
multi-Poisson one}:
\be\label{dispdeltaE}
\sqrt{\left<\delta E^2\right>}= {1 \over 2^{n \over 2}}\,E_0
\sqrt{p_1\bar{N}\frac{\alpha_1-1}{\alpha_1-3}},
\ee
where $\alpha_1$ is given in Table~1. The quadratic mean $\delta E (\delta
\phi)$ in Eq.~(\ref{dispdeltaE}) characterizes the magnitude of the disbalance
in the minijet-generated transverse energy flow.  Note that $\delta E$ is
essentially sensitive to the overall magnitude of the semihard (minijet
generated) transverse energy flow.  In Eq.~(\ref{dispdeltaE}) this is clearly
seen from $\left<\delta E^2\right> \propto E_0^2\,{\bar N}$.  Numerical values
for $\sqrt{\left<\delta E^2\right>}$ are presented in Table 4.
From now on we confine our discussion to central PbPb collisions.

\begin{figure}[p]
\vspace{-1.5cm}
 \begin{center}
 \epsfig{file=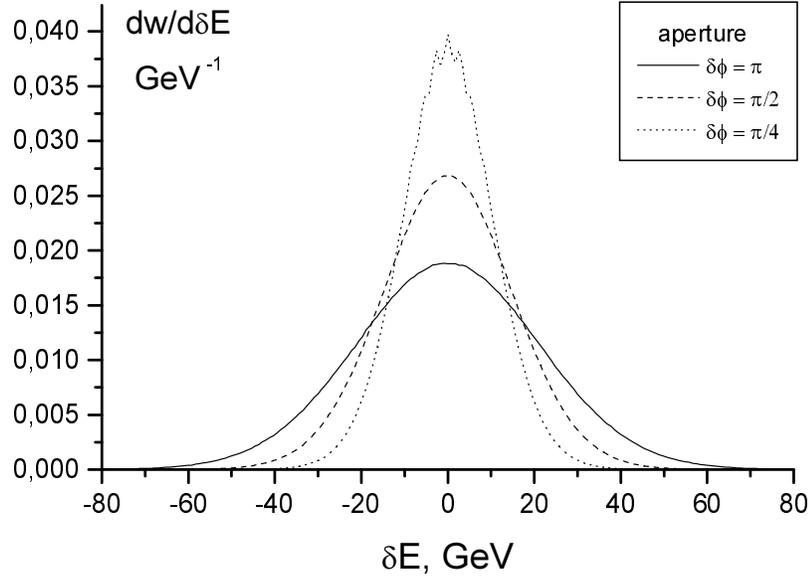,height=8cm}
 \end{center}
\vspace{-0.5cm}
 \figcaption{Probability distribution for the azimuthal asymmetry
 $\delta E_\perp$
 in unit central rapidity window at RHIC energy $\sqrt{s}=200$ GeV
 for central PbPb collisions, $\sigma^{soft}=0$}
 \label{2wo0}
\end{figure}

\begin{figure}[p]
 \begin{center}
 \epsfig{file=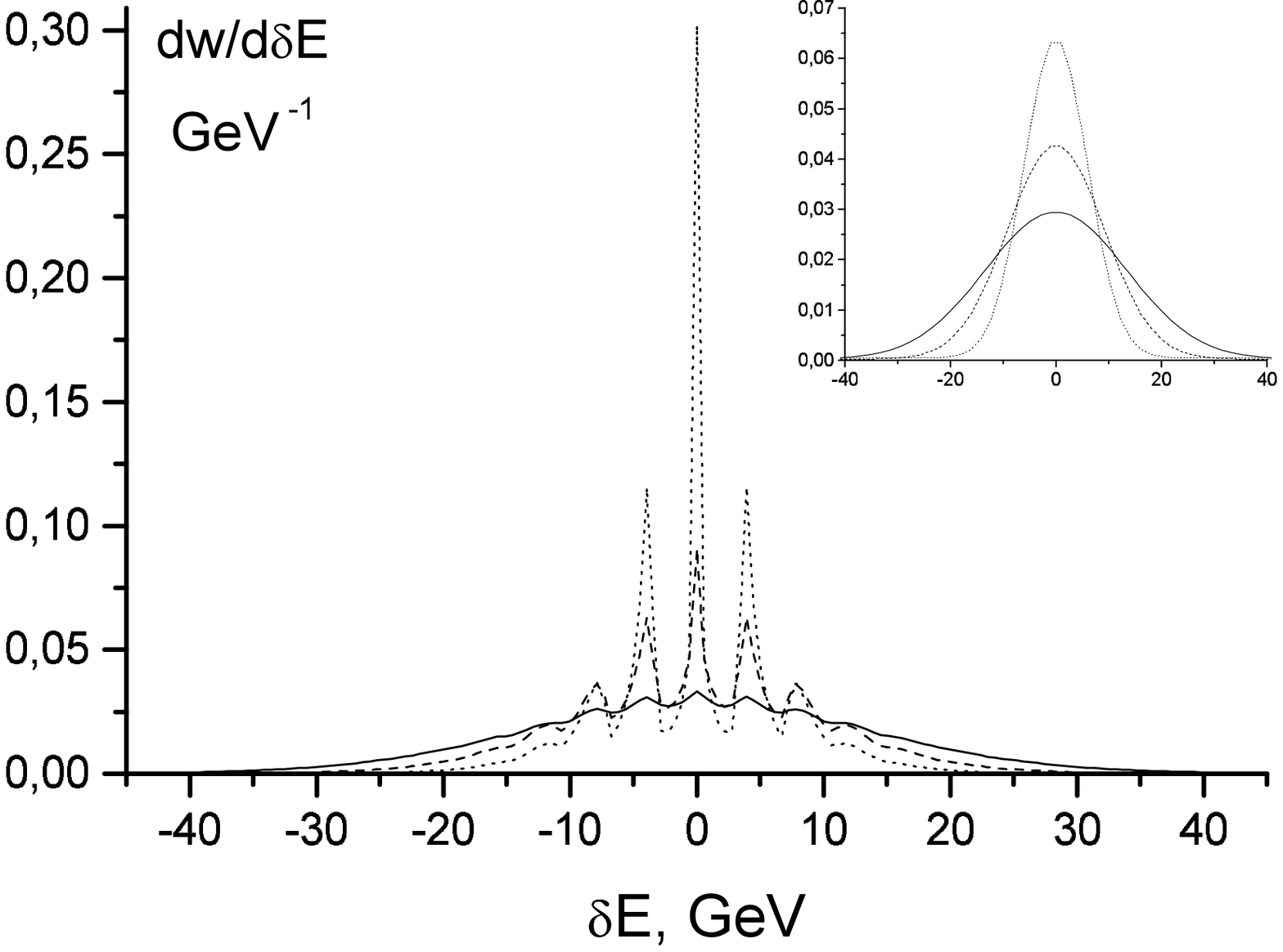,height=8cm}
 \end{center}
\vspace{-0.5cm}
 \figcaption{Probability distribution for the azimuthal asymmetry
 $\delta E_\perp$
 in unit central rapidity window at RHIC energy $\sqrt{s}=200$ GeV
 for central PbPb collisions, $\sigma^{soft}=32$\ mb}
 \label{2w0}
\end{figure}

\begin{figure}[p]
 \begin{center}
 \epsfig{file=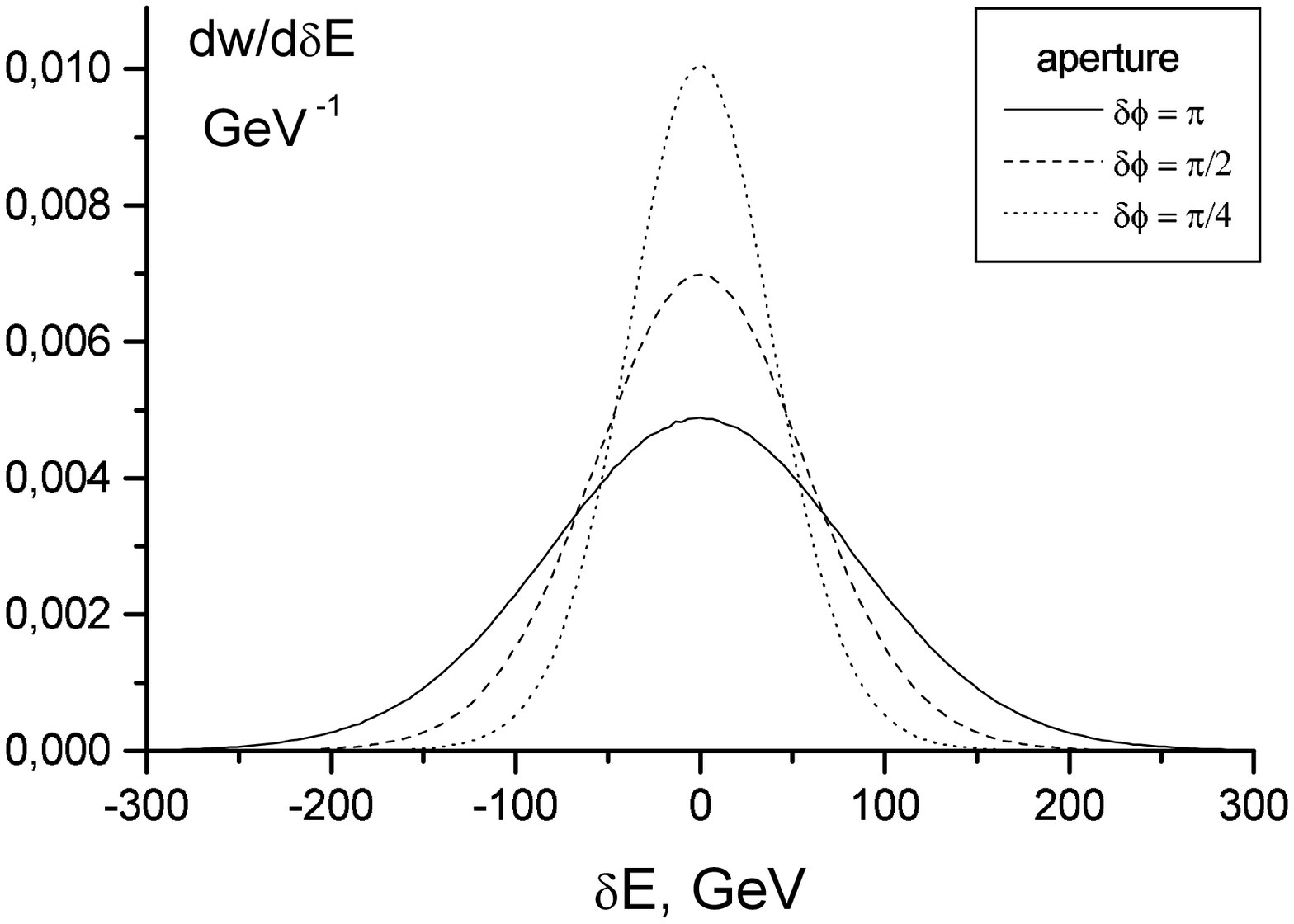,height=8cm}
 \end{center}
\vspace{-0.5cm}
 \figcaption{Probability distribution for the azimuthal asymmetry
 $\delta E_\perp$
 in unit central rapidity window at LHC energy $\sqrt{s}=5.5$ TeV
 for central PbPb collisions, $\sigma^{soft}=0$}
 \label{55wo0}
\end{figure}

\begin{figure}[p]
 \begin{center}
 \epsfig{file=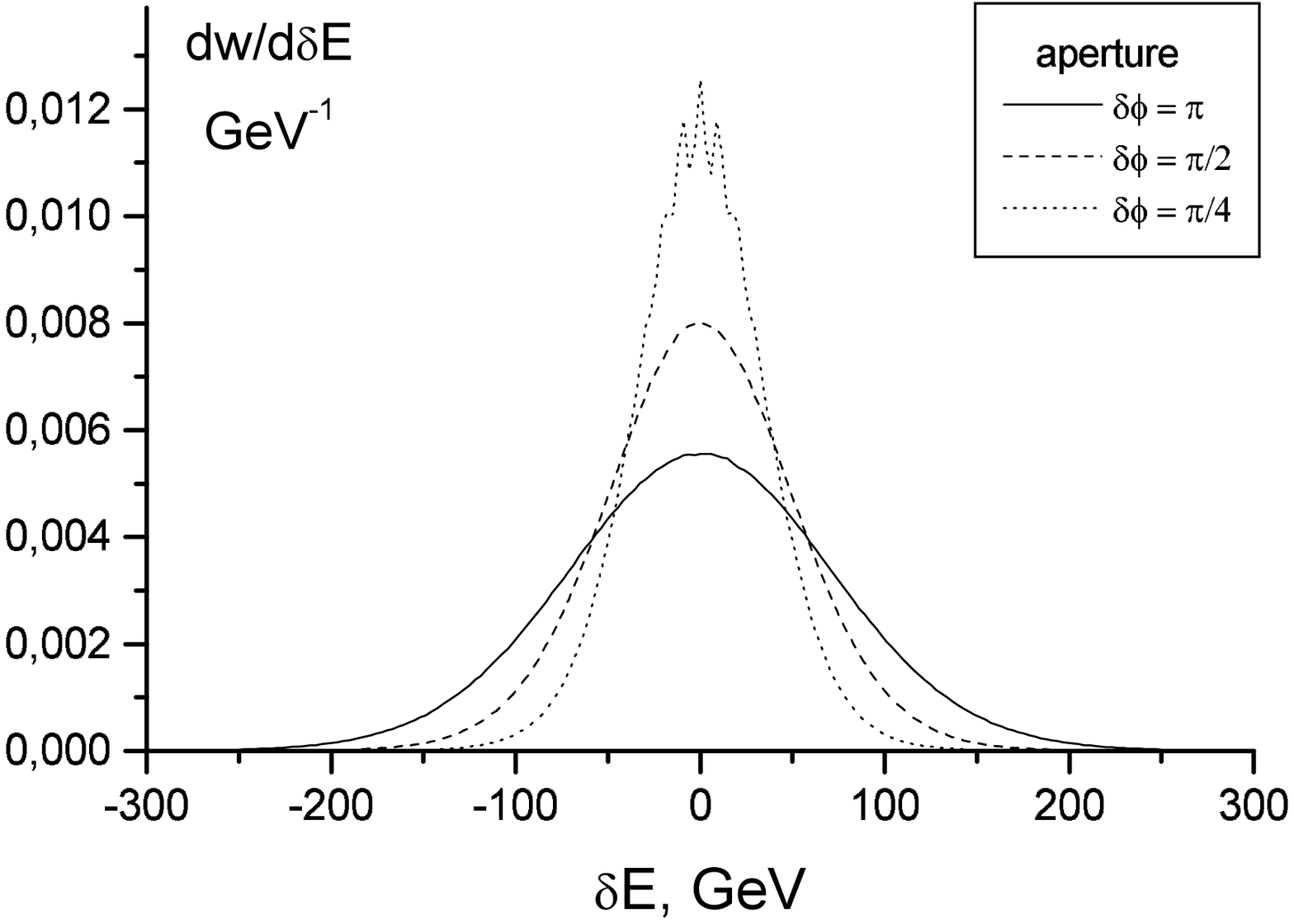,height=8cm}
 \end{center}
\vspace{-0.5cm}
 \figcaption{Probability distribution for the azimuthal asymmetry
 $\delta E_\perp$
 in unit central rapidity window at LHC energy $\sqrt{s}=5.5$ TeV
 for central PbPb collisions, $\sigma^{soft}=32$\ mb.}
 \label{55w0}
\end{figure}

In Figs.(\ref{2wo0})-(\ref{55w0}) we show the probability distribution
for $\delta E$ in central
PbPb collisions for two values of CMS energy $200$ GeV and $5.5$ TeV
and two choices for $E_0$ corresponding to mixed and hard scenarios.
The angular apertures were chosen to be $\pi$, $\pi/2$ and
$\pi/4$. From these figures we see that for all types of collisions
(except for Fig. (\ref{55wo0}))
there appear peaks in probability distribution at values $\delta E=nE_0$. This
is a reflection of a sharp cutoff adopted in the model and rapid decrease
of minijet cross section with increasing $E_\perp$. In the majority of cases
this effect is seen only for small values of angular opening. The crucial
parameter related to the peaks appearance is in fact $\bar{N}$. The smaller is
$\bar{N}$ the more evident are peaks. One could expect that hadronization and
soft processes accompanying minijet production smoothed away these peaks.
Curves that are initially smooth undergoes Gaussian law with dispertion
$\left<\delta E^2\right>$. Therefore, we can imagine appearance of curves
with peaks after smoothing as Gaussian with dispersions given by
Eq.~(\ref{dispdeltaE}). For Fig.~(\ref{2w0}) the result of
such smoothing is shown in the inserted plot.

Another useful
quantity is a normalized asymmetry which is, on the contrary, insensitive to
the absolute magnitude of transverse energy flow:
\be
r(\delta \varphi) \, = \,
{E_\uparrow (\delta \varphi) - E_\downarrow (\delta \varphi) \over
E_\uparrow (\delta \varphi) + E_\downarrow (\delta \varphi)},
\label{r}
\ee
where $r\in [-1,1]$. In particular, the normalized asymmetry $r$ simplifies
comparing the asymmetries at different CMS energies.  The values of $r(\delta
\varphi)$ in different collisions are  characterized by the normalized
probability distribution
\be\label{p(r)}
p(r)|_{\delta \varphi} \, = \,
\left.\frac{1}{\sigma}\frac{d\sigma}{dr}\right|_{\delta \varphi}.
\ee

To evaluate $p(r)$ we use a Monte-Carlo simulation of the process of nuclear
scattering as described above for the generated ensemble of $10^7$ PbPb
collisions at RHIC and at LHC energies. We have calculated the asymmetry
distributions $p(r)$ for the central (zero impact parameter $b=0$) collisions
and cone apertures $\pi,
\pi/2$ and $\pi/4$. The resulting probability distributions  are illustrated in
Figs.~(\ref{200_00})-(\ref{5500_032}) for mixed  ($\sigma^{soft} = 32$\ mb)
and hard ($\sigma^{soft} =0$) scenarios at RHIC and LHC energies:

\begin{figure}[p]
\vspace{-1.5cm}
 \begin{center}
 \epsfig{file=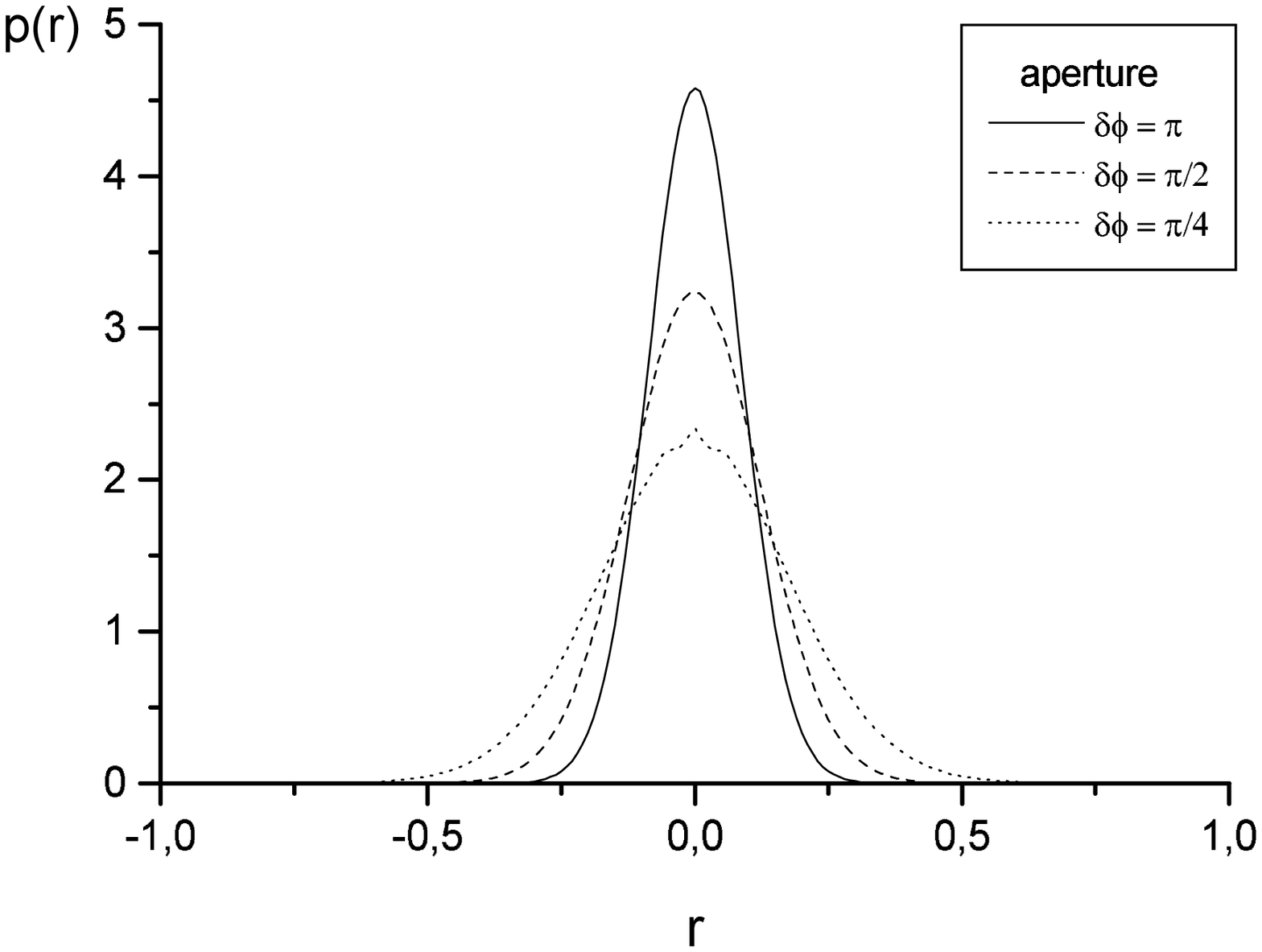,height=8cm}
 \end{center}
\vspace{-0.5cm}
 \figcaption{Probability distribution of the normalized azimuthal
 asymmetry $p(r)$
 in unit central rapidity window at RHIC energy $\sqrt{s}=200$ GeV
 for central PbPb collisions, $\sigma^{soft}=0$}
 \label{200_00}
\end{figure}

\begin{figure}[p]
 \begin{center}
 \epsfig{file=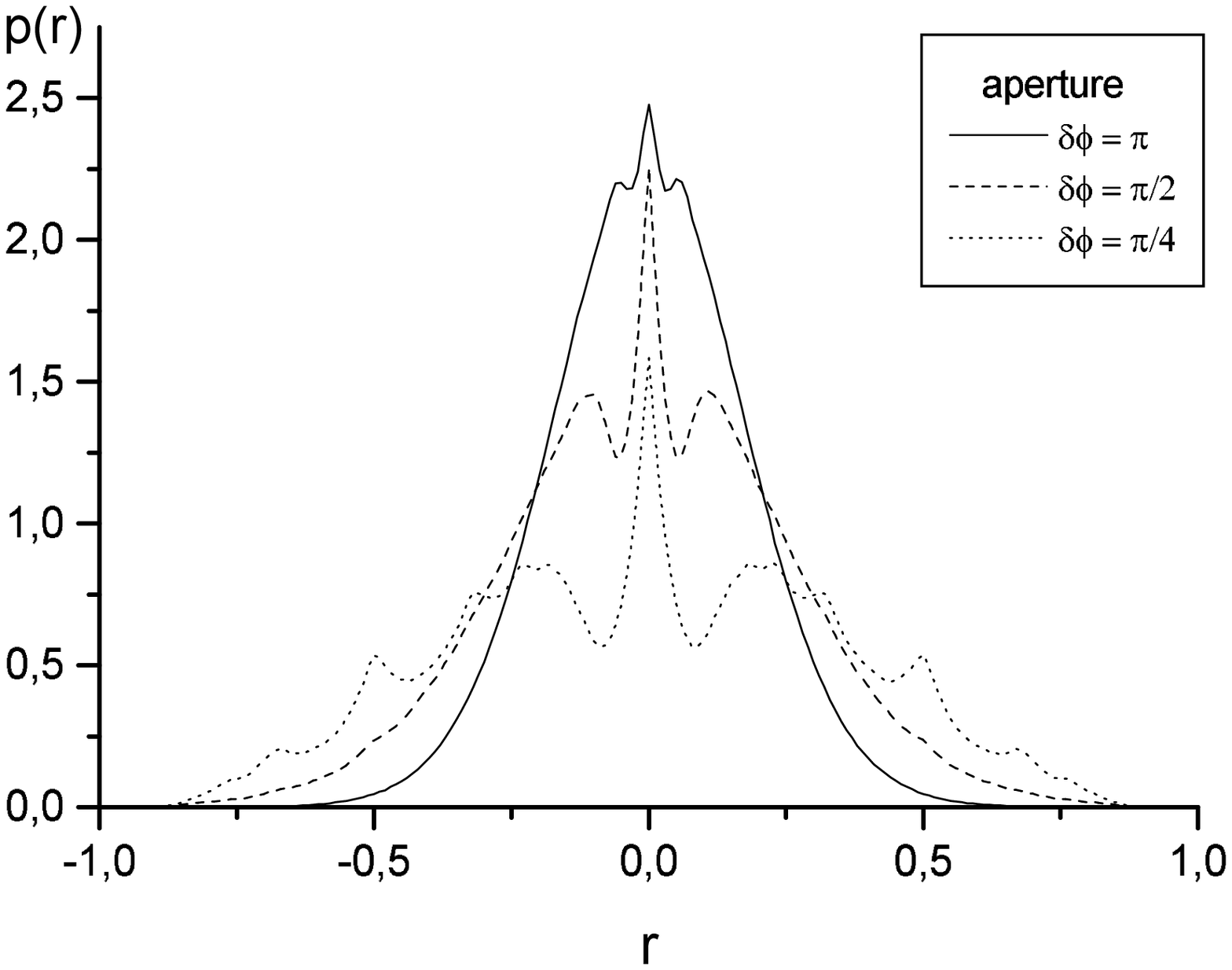,height=8cm}
 \end{center}
\vspace{-0.5cm}
 \figcaption{Probability distribution of the normalized azimuthal
 asymmetry $p(r)$
 in unit central rapidity window at RHIC energy $\sqrt{s}=200$ GeV
 for central PbPb collisions, $\sigma^{soft}=32$\ mb}
 \label{200_032}
\end{figure}

\begin{figure}[p]
 \begin{center}
 \epsfig{file=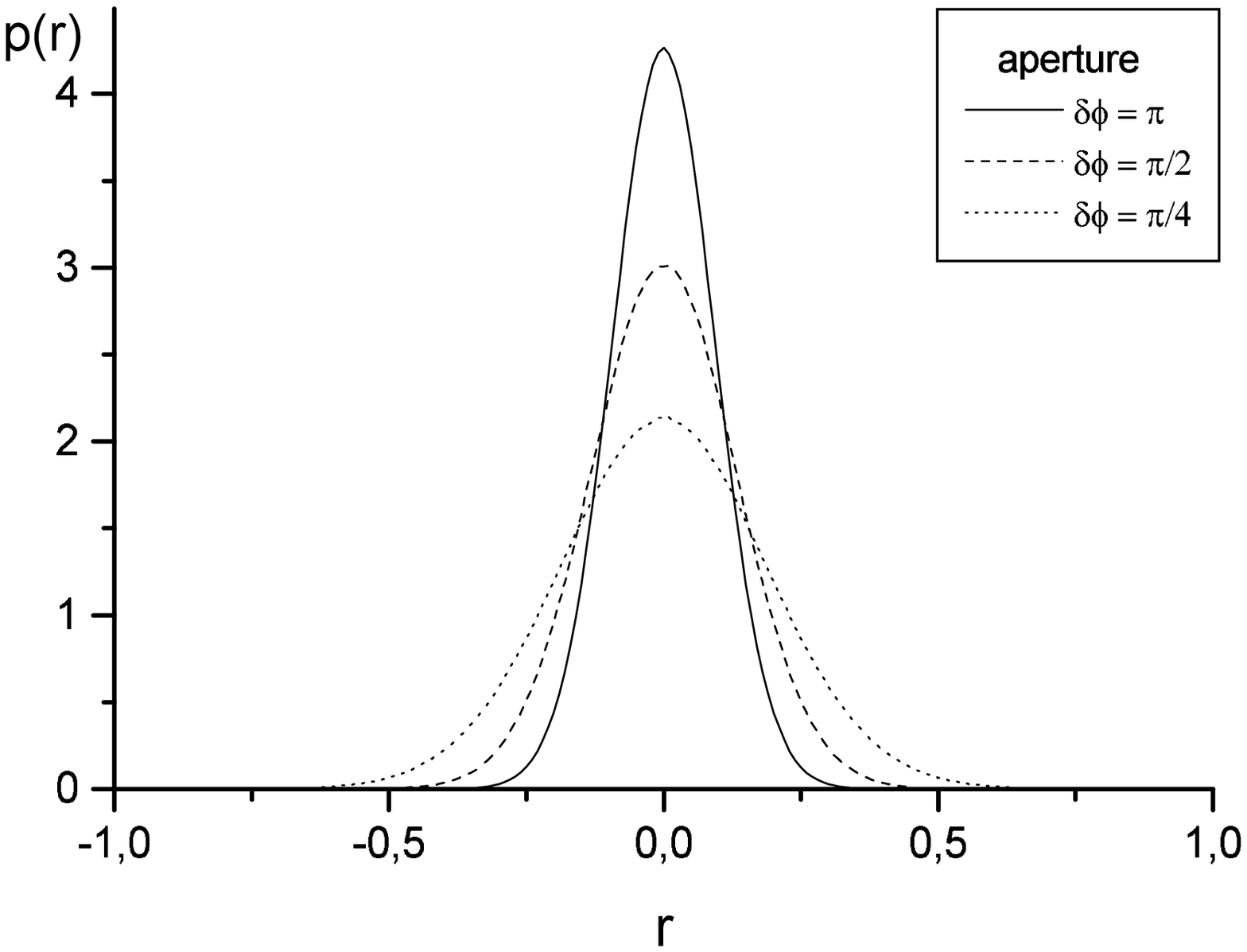,height=8cm}
 \end{center}
\vspace{-0.5cm}
 \figcaption{Probability distribution of the normalized
 azimuthal asymmetry $p(r)$
 in unit central rapidity window at LHC energy $\sqrt{s}=5.5$ TeV
 for central PbPb collisions, $\sigma^{soft}=0$}
 \label{5500_00}
\end{figure}

\begin{figure}[p]
 \begin{center}
 \epsfig{file=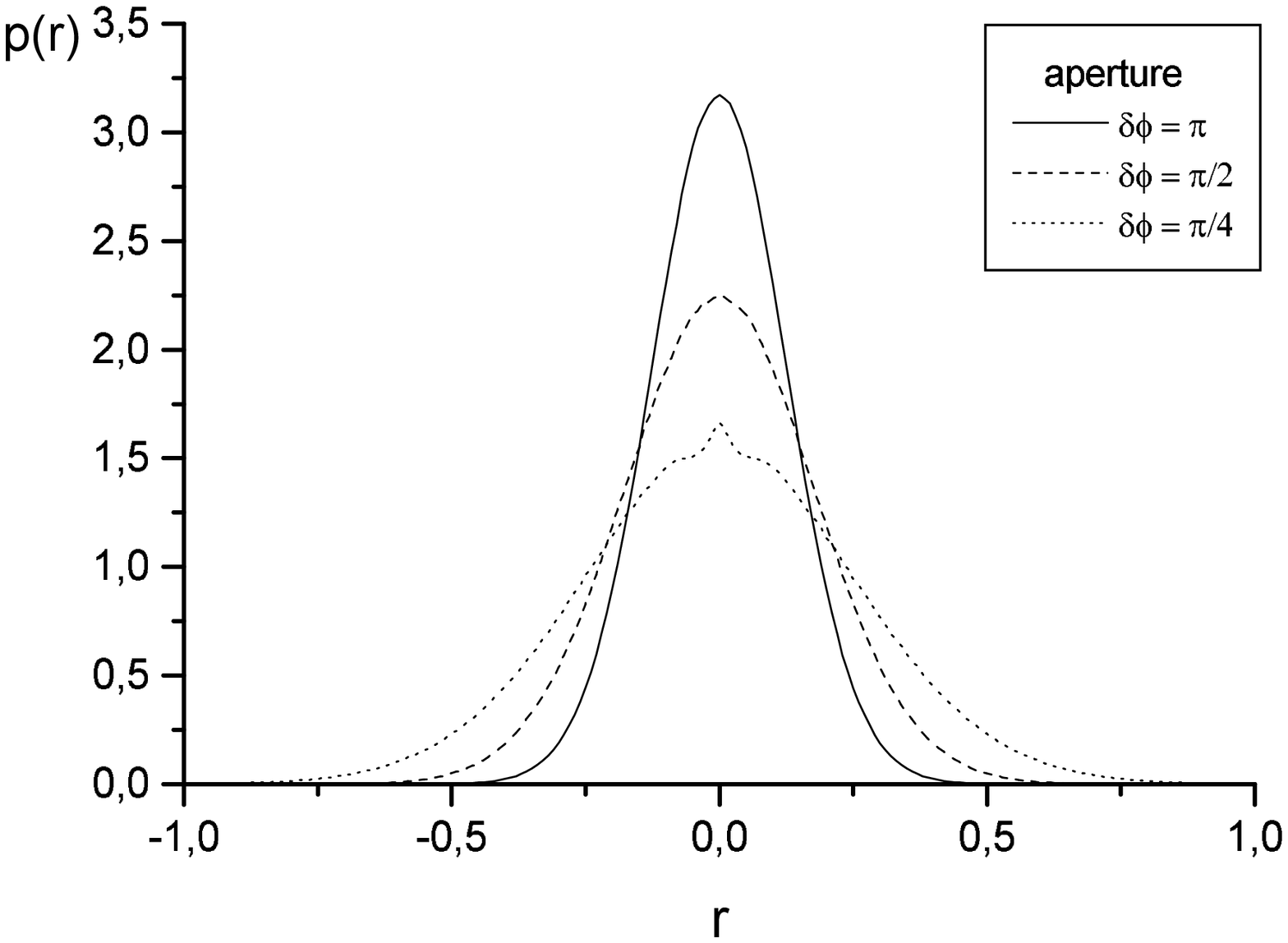,height=8cm}
 \end{center}
\vspace{-0.5cm}
 \figcaption{Probability distribution of the normalized azimuthal
 asymmetry $p(r)$
 in unit central rapidity window at LHC energy $\sqrt{s}=5.5$ TeV
 for central PbPb collisions, $\sigma^{soft}=32$\ mb.}
 \label{5500_032}
\end{figure}

Let us note that, in particular in the cases where the number of contributing
collisions is not large (RHIC), one encounters  "singular" configurations, for
which  $r=-1,1$ and $r=0$ corresponding to absolutely asymmetric and absolutely
symmetric events in PbPb collisions.  These are the events in which only one
one-jet event contributes to the given aperture during the collision ($r=-1,1$)
or one two-jet event contributes to $r=0$.  Their probabilistic weight can be
described by $\delta$ - functional contribution to $p(r)$ at the "singular"
points. Their yield in the minijet event ensemble is
given in Table~3 for mixed
($\sigma^{soft}=32$\,mb) scenario at RHIC energy (other values are
negligible).

\begin{center}
\begin{tabular}{|c|c|c|c|}
\hline
$\delta \varphi$ & $r=-1$   & $r=0$ & $r=1$ \\
\hline
$\pi/2$ & $1.3\cdot 10^{-3}$ & $1.6\cdot 10^{-2}$ & $1.3\cdot 10^{-3}$\\
$\pi/4$ & $2.5\cdot 10^{-2}$ & $1.2\cdot 10^{-1}$ & $2.5\cdot 10^{-2}$\\
\hline
\end{tabular}

\bigskip

{\bf Table~3}
 \end{center}

In Fig.~(\ref{200_032}) these contributions would correspond to infinitely
narrow peaks and are thus not shown.

The angular pattern of the transverse energy production as characterized by the
considered energy-energy azimuthal correlation probability distribution is
conveniently described by the lowest moments of $p(r)$. In Table~4 we present,
together with the numerical values of the quadratic mean $\sqrt{\left<\delta
E^2\right>}$, cf. Eqs.~(\ref{deltaE}) and (\ref{dispdeltaE}), the values of the
standard deviation $a$ defined as
\be\label{stdev}
a^2\, = \, \int dr\, (r-{\bar r})^2\, p(r),
\ee
where in our case $\left< r\right>=0$:

\medskip

\begin{center}
\begin{tabular}{|c|c|r|c|c|c|c|c|}
\hline
$\sigma^{soft}$
& {$\sqrt{S}$} &
\halfs{$\bar{N}_{PbPb}$} & $\sqrt{\left<\delta E^2\right>}$&
 \halfs{$\sqrt{p_1/\bar{N}}$}&\multicolumn{3}{|c|}{$a$}\\
\cline{6-8}
mb&GeV&&GeV&&{$\delta\phi=\pi$}
& {$\delta\phi=\pi/2$}& {$\delta\phi=\pi/4$}\\
\hline
\halfs{0}&  200 & 75.7 & 21& 0.084 & 0.088  & 0.124& 0.178\\
\cline{2-8}
& 5500 & 87.4 & 84& 0.086 & 0.094  & 0.133& 0.189\\
\cline{2-8}
\hline
\halfs{32}&  200 & 17.1 & 14& 0.168 & 0.177  & 0.259& 0.387\\
\cline{2-8}
&5500 & 47.1 & 74& 0.117 & 0.127  & 0.180& 0.257\\
\cline{2-8}
\hline
\end{tabular}

\bigskip
{\bf Table~4 }
\medskip

\end{center}

\medskip
Note that all the data presented in Table~4 include contributions from the
singular points $r=-1,0,1$.

Let us now turn to the analysis of the results presented in Figs.~(4)-(11) and
Tables~3-4. The main goal is to understand the dependence of the angular
pattern of the transverse energy flow on the basic parameters such as the
infrared cutoff $E_0$, total number of minijet-generating collisions ${\bar
N}$, the yield of asymmetric $pp$ collisions $p_1$ and CMS energy $\sqrt{s}$.

In the forth column in Table~4 we show the quadratic mean $\left<\delta
E^2\right>$ for the azimuthal opening $\delta \phi = \pi$. The results agree
with Eq.~(\ref{dispdeltaE}), so that the average disbalance in the
transverse energy is indeed essentially determined by $E_0$ and ${\bar N}$.

To understand the results
for normalized asymmetry $p(r)$
it is helpful to consider a simplified model, in
which the elementary $pp$ collisions can produce only some given amount of
transverse energy,
\be\label{desing}
\left ( {d \sigma \over d E_{\perp}} \right )^{pp} \, = \,
\sigma^{hard}(\sqrt{s})\, \delta(E_{\perp}-E_0(\sqrt{s})),
\ee
so that all transverse energy is assumed to be produced exactly at the cutoff $E_0$. Note that except for ascribing the
energy production to elementary $pp$ collisions this model is very similar to
the expected pattern of transverse energy production in the quasiclassical
approach based on McLerran-Venugopalan model, cf. \cite{AM}. Then, for
considered azimuthal apertures $\delta \phi = \pi/2^n$\,\,$(n=0,1,2)$ we get
for the standard deviation $a$ defined in Eq.~(\ref{stdev})
\footnote{The details of this calculation can be found in Appendix A}:
\be\label{amod}
a \left ( \delta \phi = {\pi \over 2^n} \right ) \,
\approx
\, \sqrt{{1 \over 2^n}\,  {p_1 \over {\bar N}}}
\left (
 1 +  O \left ( {1 \over {\bar N}} \right )
\right )
\ee
This shows that the width of the
distribution $p(r)$ is determined by the ratio of the relative yield of
asymmetric collisions $p_1$ to the average number of collisions. In Table~4 we
compare the predictions of this simple model to the values of standard deviation
$a$ computed using the differential spectra plotted in Figs~(1),~(2) (to save
space, only the results for $\delta \phi=\pi$ are given in column 5) and
observe only a 10 \% difference. This shows that the results obtained using the
continuous spectra in Figs.~(1),~(2) are essentially determined by the
contribution at the cutoff energy $E_0$.

From Fig.~(\ref{200_032})
we see that for small number of asymmetric collisions the
shape of $p(r)$ has peculiar sharp peaks at certain values of $r$. The origin
for this is in fact the growth of the differential cross section for transverse
energy production at small $E_{\perp}$ in \pp collisions, cf. Figs.~ 1-2.
Indeed, let us, for simplicity, assume that each \pp collision in the restricted
minijet ensemble can produce the transverse energy exactly at the cutoff
$E_\perp=E_0$ only, cf. Eq.~(\ref{desing}).  In this case in addition to the
 "true" singular points $r=-1,0,1$ we will have "semisingular" ones so that for
 a particular event containing $n$ minijets, with $n_{1\uparrow}$ being the
number of "up-coming" one-jet events, $n_{1\downarrow}$ being the number of
"down-coming" one-jet events, and $n_2=n-n_{1\uparrow}-n_{1\downarrow}$ being
the number of two-jet events, the following exact relation does hold:
\be
r=\frac{n_{1\uparrow}-n_{1\downarrow}}{n}
\ee
Thus the values of $r$ belong to a set of {\it rational} numbers in the
interval $[-1,1]$, which we call "semisingular". Of course, the most
spectacular "semisingular" points are those with small numerators and
denominators both due to the higher frequency of events having small number of
minijets and to smaller distribution width (deviation from $E_{\perp}=n\,E_0$)
for events with small number of asymmetric collisions.

Let us note that the appearance of singular points $-1, 0, 1$ is a consequence
of calculating the cross sections for transverse energy production in the
elementary hard block in the lowest order in perturbation theory. In the
next-to-leading order, where the transverse energy can be  shared between three
(mini)jets these singular  points will become milder singularities of $p(r)$ at
$r=-1, 0, 1$. This shows that the calculation of the true shape of $p(r)$ near
the singular points requires, as usual,  resumming the perturbative
contributions to all orders.

Physically, within the scheme adopted in this paper, the number of semihard
collisions depends on the sharing of inelasticity between soft and hard
mechanisms of transverse energy production. In the mixed scenario we assumed
that $32$\, mb of the inelastic cross section of $pp$ collisions corresponds to
the soft production mechanism, while the rest of inelastic cross section is due
to semihard production. In the hard scenario it is assumed that the semihard
transverse energy production saturates all available inelasticity. It is to be
expected that the distributions characterizing azimuthal asymmetry of
transverse energy production $p(r)$ defined in Eq.~(\ref{p(r)}) will be wider
in the mixed scenario than in the hard one. From Table~5 we see that this is
indeed the case. At RHIC energies the standard deviation for mixed scenario is
bigger than that in hard one in the ratio $a^{mixed}_{RHIC}/a^{hard}_{RHIC}
\simeq 2.0-2.1$. At LHC energies the effect is less pronounced, here
$a^{mixed}_{LHC}/a^{hard}_{LHC} \simeq 1.3-1.4$.  We see that with growing CMS
energy the angular pattern of the minijet-generated transverse energy flow is
becoming less sensitive to the relative weight of perturbative and
nonperturbative contributions to the inelastic cross section.

The dependence of the standard deviation $a$ on the aperture, remains
essentially the same  for both CMS energies and values of the impact parameter
considered  and is inversely proportional to the angular opening:
\be
a_{\pi/2n} \, \simeq \, 2^{{n \over 2}}\, a_{\pi}
\ee
which is consistent with the prediction of a simple model of transverse energy
production Eq.~(\ref{amod}) and corresponds to purely statistical change in the
standard deviation, where shrinking the angular aperture by a factor of $2$
enlarges the standard deviation by a factor of $\sqrt{2}$.

\section{Conclusions}

The main results of our analysis can be formulated as follows.

We first discussed a basic asymmetry in the minijet transverse energy
production in a restricted rapidity window in $pp$ collisions due to different
probabilities of having a "symmetric" two-jet or "asymmetric" one-jet
contribution in the rapidity interval under consideration. The cross sections
for symmetric and asymmetric contributions in $pp$ collisions for RHIC and LHC
energy show that while at RHIC energy the weight of both configurations is
approximately equal, at LHC energy the asymmetric contribution is clearly
dominant.

We further considered a geometrical model for nuclear collisions in which they
are described as an incoherent superposition of nucleon-nucleon ones.  We
discussed two possible partitions of the inelastic cross section in terms of
 soft and semihard contributions and analyzed the angular pattern of
minijet-generated energy flow for central and peripheral nuclear collisions at
RHIC and LHC energies. Specifically we considered probability distributions for
transverse energy-transverse energy correlations in the oppositely azimuthally
oriented cones with varying aperture. We show that the resulting distributions
are essentially sensitive to the number of semihard collisions, which is in
turn dependent on the above-mentioned partition of the inelastic cross section
into contributions of different types, and on the (related) choice of the
infrared cutoff. We also show that the results are very close to the
predictions of a simple model, in which all the transverse energy is produced
directly at the infrared cutoff.

 The approach developed in this paper could be further generalized to the
analysis of the minijet generated background oriented flow \cite{LOO} (for the
definition of the oriented flow and comprehensive discussion see e.g.
\cite{JYO}). In particular, as the importance of minijet contribution is
expected to grow with energy the presence of the background oriented flow of
purely fluctuational origin could increasingly influence corresponding hadronic
observables.

 Another crucial issue is the dynamical evolution of the primordial partonic
inhomogenities in the course of parton-hadron conversion. In a recent study
\cite{WS} it was shown, that the seed inhomogenity in the initial condition of
the elliptic flow type for the hadronic RQMD code survives the freeze-out  and
is visible in final azimuthal distributions. This question is surely most
important and will be discussed in the forthcoming publication \cite{LO1}.

\begin{center}
{\it Acknowledgements}
\end{center}
\medskip

We are grateful to K.~Kajantie, G.~Zinovjev, J.~Schukraft, K.~Eskola,
I.M~Dre\-min, I.V.~Andreev, J.-P.~Blaizot, J.-Y.~Ollitrault, L.~McLerran and
P.~Jacobs for useful discussions.

A.L. is grateful for kind hospitality and support at CERN Theory Division,
where this work was started and at Service de Physique Theorique de Saclay,
where the major part of it was done.

We are grateful to the referee of the first version of this paper for
constructive remarks and suggestions.

This work was also partially supported by Russian Fund for Basic Research,
Grant 96-02-16347.

The work of D.O. was partially supported by the INTAS grant 96-0457 within the
ICFPM program.

\newcommand{\appex}{ \setcounter{equation}{0}\section}

\setcounter{section}{0}
\renewcommand{\thesection}{{\it Appendix} \Alph{section}}
\renewcommand{\theequation}{\Alph{section}.\arabic{equation}}
\renewcommand{\thesubsection}{\Alph{section}.\arabic{subsection}}

\appex{$\left<r^2\right>$ calculation}

In this Appendix we present a derivation of the formula for the standard
deviation $a$ Eq.~(\ref{amod}).

Let $n$ be the number of minijet producing hadron-hadron collision
in a given the nucleus-nucleus one characterized by the Poissonian distribution
Eq. (\ref{Poisson}). Let us further denote by $n_1$ the number of
those minijet producing hadron collisions in which only one minijet
hits the rapidity window under consideration (asymmetric contribution).  and by
$n_{1\uparrow}$ and $n_{2\downarrow}$ are the numbers of such single minijets
propagating into the upper and lower of the two oppositely oriented cones
correspondingly.
In what follows we shall use a simplified model of transverse energy production
in which it is produced strictly at the cutoff $E_0$ (cf. Eq.~(\ref{desing}).
The averaging over event ensemle has to be
done in the sequence opposite to the one adopted in the Monte Carlo procedure.
First, we  average over $n_{1\uparrow}$ at fixed
$n_1=n_{1\uparrow}+n_{1\downarrow}$:
\be
\left<r^2\right>_{n_1}=\sum\limits_{n_{1\uparrow}=0}^{n_1}
\frac{1}{2^{n_1}}C_{n_1}^{n_{1\uparrow}}
\frac{(n_{1\uparrow}-n_{1\downarrow})^2}{n^2}=\frac{n_1}{n^2}
\ee
Next, we average over $n_1$ at fixed $n$ according to the
binomial probability distribution  Eq. (\ref{binom}):
\be
\left<r^2\right>_{n}=\frac{p_1}{n}
\ee
Finally, we have to average over the Poissonian distribution
Eq. (\ref{Poisson}) yielding
\be
a^2=\left<r^2\right>_{n}=
\e^{-\bar{N}}\sum\limits_{n=1}^{\infty}\frac{p_1}{n}
\frac{\strut\bar{N}^n}{n!}=p_1 {\bar{N}}{\e^{-\bar{N}}}\
F(1,1;2,2;\strut\bar{N})
\approx\frac{p_1}{\strut\bar{N}}
\left(1+O\left(\frac{1}{\strut\bar{N}}\right)\right),
\ee
Where $F(1,1;2,2;\strut\bar{N})$ is a generalized hypergeometric function.

\end{document}